\documentclass{emulateapj}

\begin{document}

\shortauthors{Luhman}

\shorttitle{Census of Disk Populations in Sco-Cen}

\title{A Census of the Circumstellar Disk Populations in the Sco-Cen Complex\altaffilmark{1}}

\author{K. L. Luhman\altaffilmark{2,3}}

\altaffiltext{1}
{Based on observations made with the Gaia mission, the Two Micron All
Sky Survey, and the Wide-field Infrared Survey Explorer.}

\altaffiltext{2}{Department of Astronomy and Astrophysics,
The Pennsylvania State University, University Park, PA 16802, USA;
kll207@psu.edu}
\altaffiltext{3}{Center for Exoplanets and Habitable Worlds, 
The Pennsylvania State University, University Park, PA 16802, USA}

\begin{abstract}

I have used mid-infrared (IR) photometry from the Wide-field Infrared Survey
Explorer (WISE) to perform a census of circumstellar disks among $\sim$10,000
candidate members of the Sco-Cen complex that were recently identified
with data from the Gaia mission.
IR excesses are detected for more than 1200 of the WISE counterparts that are
within the commonly adopted boundary for Sco-Cen, $\sim400$ of which
are newly reported in this work. 
The richest population in Sco-Cen, UCL/LCC, contains the largest available
sample of disks ($>$500) for any population near its age 
($\sim$20~Myr). UCL/LCC also provides the
tightest statistical constraints to date on the disk fractions of low-mass
stars for any single age beyond that of Upper Sco ($\sim11$~Myr).
For Upper Sco and UCL/LCC, I have measured the disk fractions
as a function of spectral type.
The disk fraction in Upper Sco is higher at later spectral types, which is 
consistent with the results for previous samples of candidate members.
In UCL/LCC, that trend has become more pronounced;
the disk fractions in UCL/LCC are lower than those in Upper Sco by factors
of $\sim$10, 5.7, and 2.5 at B7--K5.5, K6--M3.5, and M3.75--M6, respectively.
The data in UCL/LCC also demonstrate that the disk fraction for low-mass stars
remains non-negligible at an age of 20~Myr (0.09$\pm$0.01).
Finally, I find no significant differences in the ages of disk-bearing
and diskless low-mass stars in Upper Sco and UCL/LCC based on their
positions in color-magnitude diagrams.

\end{abstract}

\section{Introduction}
\label{sec:intro}

The lifetimes of primordial circumstellar disks around newborn stars
constrain the time available for the formation of giant planets
\citep{pol96,bos98}. Those lifetimes are estimated by
comparing the prevalence of disks among clusters and associations that
span a range of ages
\citep{hai01,mam09,clo14,rib15,men17,ric18}\footnote{Disk lifetimes 
are subject to the systematic errors in age estimates for young
stellar populations \citep{may08,bel13,her15}.}.
The stars that harbor disks in a given population can be identified using
either accretion signatures or infrared (IR) excess emission from cool dust,
where the latter is usually easier to measure for large samples of stars.
To detect disks via IR excesses, the imaging should be performed at wavelengths
that are long enough that typical disks dominate the stellar photospheres
(i.e., excesses are large) but short enough that the sensitivity
reaches photospheric flux levels,
which corresponds to $\sim$4--10~\micron.
Observations at longer IR wavelengths are also useful for detecting more 
evolved disks that have developed inner holes and thus lack the hot dust
that produces excesses at $<10$~\micron.

The Spitzer Space Telescope \citep{faz04,rie04,wer04} and the Wide-field 
Infrared Survey Explorer \citep[WISE,][]{wri10} have been the best available 
facilities for detecting disks via IR excesses. Spitzer offered higher 
sensitivity and angular resolution while WISE provided data for the entire sky, 
making them well-suited for compact star-forming clusters
\cite[e.g.,][]{lad06,sic06,cie07,cur07,her07,luh08} and widely-distributed
associations \citep{riz12}, respectively.
Based on data from Spitzer and WISE, disk fractions decrease significantly
between ages of 1--10~Myr with low-mass stars experiencing
the slowest decline \citep{car06,ken09,rib15}.
A substantial fraction of low-mass stars retain disks at ages of $\sim10$~Myr
\citep[$\sim20$\%,][]{luh20u} and a few at ages of $\sim$20--50~Myr
have been found to harbor disks that may be primordial
\citep{whi05,har05,bou16,sil16,sil20,mur18,lee20}\footnote{It is unclear
whether these disks should be classified as old primordial disks or young 
debris disks \citep{fla19}.}.
Disk fractions for low-mass stars are uncertain at ages beyond
10~Myr because most clusters and associations at those ages
are too distant for available mid-IR data to reach $\sim$0.1~$M_\odot$ or
have insufficient numbers of known members for good statistical constraints.

The oldest populations in the Scorpius-Centaurus complex 
\citep[Sco-Cen,][]{pm08} contain several thousand stars with ages
of $\sim$20~Myr and distances of 100--200~pc \citep{dam19,luh20u},
making Sco-Cen the most promising site for measuring a disk fraction
for low-mass stars at an age greater than 10~Myr.
The primary obstacle in capitalizing on Sco-Cen for that purpose
has been the difficulty in identifying its members, which are distributed
across a large area of sky ($\sim60\arcdeg\times20\arcdeg$).
The high-precision astrometry and photometry from the Gaia mission
\citep{per01,deb12,gaia16b} have made it possible to perform a thorough
census of Sco-Cen down to the hydrogen burning mass limit.
Several studies have used Gaia data to identify candidate members of
populations within Sco-Cen \citep[e.g.,][]{dam19,gal20} and some of
those candidates have been examined for evidence of disks 
\citep[e.g.,][]{gol18,tei20}.
I have used the early installment of the third data release of Gaia
\citep[EDR3,][]{bro21}
to perform a census of the stellar populations in the Sco-Cen complex
\citep{luh21}. In this paper, I present a census of the circumstellar
disks among those candidate members based on mid-IR photometry from WISE.

\section{Identification of Disks in Sco-Cen}

\subsection{Gaia Census of Stellar Members of Sco-Cen}
\label{sec:sample}

Early studies of the Sco-Cen OB association divided it into three 
subgroups that are adjacent on the sky: Upper Sco, Upper Centaurus-Lupus (UCL),
and Lower Centaurus-Crux \citep[LCC,][]{bla64,dez99}.
UCL and LCC are now known to form a single continuous distribution of stars
that extends across Upper Sco \citep{riz11,luh21}.
Age estimates for Upper Sco and UCL/LCC have ranged from 5--12~Myr and 
15--21~Myr, respectively 
\citep{deg89,pre02,pec12,son12,her15,pec16,dav19b,luh20u,luh21},
where the older values are favored by the more recent work.
Gaia data have revealed an additional group associated with the star
V1062~Sco that is coeval with UCL/LCC \citep{ros18,dam19}.
These populations contain several thousand stars and overlap spatially and
kinematically with smaller groups of younger stars ($\lesssim6$~Myr)
associated with the Ophiuchus and Lupus clouds.

In \citet{luh21}, I performed a survey for members of the populations
in the Sco-Cen complex -- Upper Sco, UCL/LCC, the V1062~Sco group,
Ophiuchus, and Lupus -- using astrometry and photometry from Gaia 
EDR3\footnote{The photometry from Gaia is in bands at 3300--10500~\AA\ ($G$), 
3300--6800~\AA\ ($G_{\rm BP}$), and 6300-10500~\AA\ ($G_{\rm RP}$).}.
That analysis identified 10,509 candidate members of Sco-Cen with parallax
errors of $\sigma_{\pi}<1$~mas and 195 additional sources
that did not satisfy the kinematic criteria for membership but that are
possible companions to candidates in the first sample
based on their small separations ($<5\arcsec$) and
positions in color-magnitude diagrams (CMDs), IR excesses, or roughly
similar proper motions and parallaxes.
Spectral classifications were available for more than 3000 candidates,
and extinctions were derived for those stars from optical and near-IR colors 
(e.g., $G_{\rm RP}-J$) relative to the typical intrinsic values for young stars.
For the remaining candidates, both spectral types and extinctions were
estimated by dereddening their observed colors to the sequences
of intrinsic colors in color-color diagrams.

\subsection{Compilation of Infrared Photometry}

WISE is the only available source of mid-IR photometry for most of the
candidate members of Sco-Cen from \citet{luh21}. As discussed in that study,
all candidates that are not blended with other stars should be detected by WISE.
The Spitzer Space Telescope provided higher sensitivity and angular resolution
than WISE, but only a small fraction of the Sco-Cen candidates are within
fields imaged by that facility. 
Most of those stars have been searched for IR excesses from disks
with the Spitzer data \citep[e.g.,][]{car06,cie07,mer08}.

The images from WISE were obtained in bands centered at 3.4, 4.6, 12, and
22~$\mu$m, which are denoted as W1, W2, W3, and W4, respectively.
I have made use of the data from the WISE All-Sky Source Catalog \citep{cut12a},
the AllWISE Source Catalog, and the AllWISE Reject Table \citep{cut13}.
AllWISE is newer than the All-Sky catalog and is more reliable than the Reject 
Table, so AllWISE is normally preferred, but some stars only have entries in
All-Sky or the Reject Table.
To facilitate the measurement of mid-IR excesses in the WISE bands, I also have
employed near-IR photometry in $J$, $H$, and $K_s$ (1.25, 1.65, 2.16~\micron)
from the Point Source Catalog of the Two Micron All Sky Survey
\citep[2MASS,][]{skr06}.

The point spread functions in the images from 2MASS and AllWISE exhibit 
FWHM$\sim$2.5--3$\arcsec$ and 6--12$\arcsec$ (W1--W4),
respectively \citep{skr06,wri10}.
The diffraction limit for Gaia is $\sim0\farcs1$ and the effective resolution
of EDR3 is $\sim0\farcs2$ \citep{fab21}.
Because of its much higher resolution, Gaia frequently resolves groups of
closely spaced stars that appear as single unresolved sources in 2MASS and WISE.

The candidate members of Sco-Cen from \citet{luh21} were selected from
Gaia EDR3. For each candidate, I calculated its expected position
at an epoch of 2010.5 (near the midpoint of the WISE survey) based on the
Gaia astrometry and identified the closest matching source within $3\arcsec$
in each of the three WISE catalogs mentioned previously. 
I selected $3\arcsec$ as the matching threshold because it should be large
enough to capture nearly all true matches. A smaller threshold for a more
reliable sample of matches can be applied in later analysis as
necessary (e.g., Section~\ref{sec:adopt}). 
I preferred the matches from AllWISE even if matches from the other
two catalogs were somewhat closer.
Therefore, I adopted the match from AllWISE unless one of the other
two catalogs offered a match that was closer to the Gaia source by
at least $0\farcs5$, in which case I adopted that match.
A threshold of $0\farcs5$ was adopted because most
matching sources between AllWISE and All-Sky have separations less than
that value. In other words, if AllWISE and All-Sky sources are separated
by $<0\farcs5$, they are likely the same object, whereas a larger
separation indicates that they may correspond to different objects.
The adopted WISE match for each Gaia source is provided in the tables of
candidates from \citet{luh21}. Those tables also include the separation
between each Gaia source and its WISE match (at 2010.5) and a flag
indicating whether the Gaia source is the closest match in EDR3 for the
WISE source. The latter flag is useful because it is common for multiple
Gaia sources to have the same WISE source as their closest match.
Whereas \citet{luh21} presents tables of Gaia sources (and the Gaia data)
and the names of matching sources in WISE (and 2MASS), Table~\ref{tab:phot} 
compiles all of those WISE matches and their data from WISE.
The tables from \citet{luh21} contain a total of 10,704 Gaia sources
while Table~\ref{tab:phot} contains 9675 WISE sources.
The latter number is smaller because a given WISE source can be matched to
multiple Gaia sources and some Gaia sources lack WISE counterparts
within $3\arcsec$, which is usually due to blending with other stars.
For each WISE source, Table~\ref{tab:phot} also includes the name
of the closest match in 2MASS within $3\arcsec$, the separation between
the 2MASS and WISE sources, and the $JHK_s$ photometry from 2MASS.

Spurious detections in the WISE catalogs are increasingly common in the
bands at longer wavelengths. Therefore, I visually inspected the AllWISE Atlas
images of all WISE matches to candidate members of Sco-Cen to check for
false detections. The latter are indicated by a flag in Table~\ref{tab:phot}. 
If a photometric measurement is flagged as a false 
detection, it has been omitted from Table~\ref{tab:phot} and is not used in 
this work. Since the WISE images are less sensitive to stellar photospheres at 
longer wavelengths, all WISE sources in Table~\ref{tab:phot} have detections 
in W1 and the fraction with detections decreases from W2 through W4.

While inspecting the WISE images, I identified 21 sources whose centroids
shifted significantly ($\gtrsim2\arcsec$) between W1 and W3/W4. 
In each case, the four bands of data are associated with a single WISE source,
but the detections at shorter and longer wavelengths likely correspond to
different objects. Most of these blended pairs are resolved by Gaia EDR3.
In Table~\ref{tab:pairs}, I present the names of the WISE sources and
the names of the Gaia sources that likely dominate in W1 and W3/W4.
I also indicate whether each pair could be a binary system based on 
a comparison of the parallaxes and proper motions of the components.
Most of the pairs are possible binary systems. In those systems, the dominance 
of the secondaries in W3 or W4 indicates that they likely harbor disks.
Two of the objects that are responsible for the W3/W4 emission are classified 
as field stars based on the Gaia data.
Two of the sources that dominate at W3/W4 are not detected by Gaia,
which may be disk-bearing companions with very low masses or 
unrelated objects (e.g., galaxies).
If the source dominating in W3/W4 is not a candidate member of Sco-Cen
from \citet{luh21} that is matched to the WISE source in that study, 
then the W3 and W4 detections are flagged as false in Table~\ref{tab:phot}.

\subsection{Measurement of Infrared Excesses}
\label{sec:exc}

I have searched for evidence of disks among the WISE sources in 
Table~\ref{tab:phot} by checking for emission in the WISE bands
in excess above that expected from stellar photospheres.
I have measured excesses in W2, W3, and W4 using colors between those
bands and W1.
The $K_s$ band is frequently used as an alternative to W1 when measuring
excesses in mid-IR bands. $K_s$ has the advantage of producing
somewhat larger color excesses while W1 avoids errors in excesses due
to variability and is better matched to the other WISE bands in angular
resolution. The two bands produce very similar results when identifying
excesses in W2--W4.

\begin{figure}
\epsscale{1.2}
\plotone{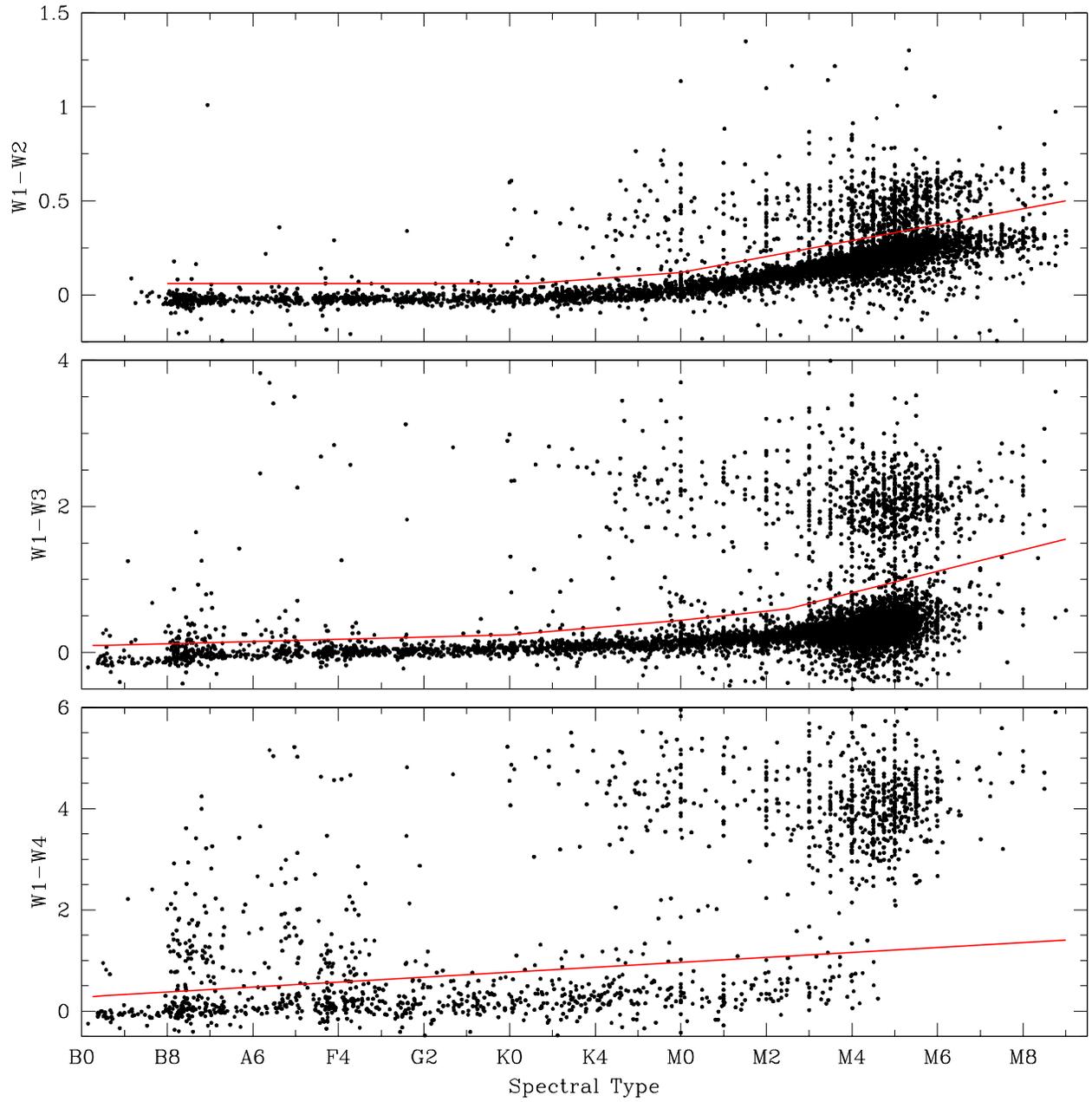}
\caption{
Extinction-corrected IR colors versus spectral type for the WISE
sources from Table~\ref{tab:phot}.
For stars that lack spectroscopy, spectral types have been estimated from
photometry \citep{luh21}.
In each diagram, the tight sequence of blue colors corresponds to stellar
photospheres. The thresholds used for identifying color excesses from
disks are indicated (red solid lines).
}
\label{fig:exc}
\end{figure}

In Figure~\ref{fig:exc}, I have plotted extinction-corrected W1$-$W2, W1$-$W3,
and W1$-$W4 versus spectral type for the WISE sources from Table~\ref{tab:phot}
that are within $1\arcsec$ of a Gaia candidate from \citet{luh21}.
The W2 data at W2$<$6 have been omitted since they are subject to significant
systematic errors \citep{cut12b}.
I have adopted the spectral types and extinction estimates for the Gaia
candidates from \citet{luh21} (see Section~\ref{sec:sample}).
For each of the three WISE colors in Figure~\ref{fig:exc}, the data exhibit
a well-defined sequence of blue sources and a broader distribution of redder
sources that correspond to stellar photospheres and disk-bearing stars,
respectively. The spectral types earlier than M0 have been shifted by 
random amounts between $\pm0.5$ subclass
in Figure~\ref{fig:exc} to give each photospheric sequence a smoother 
appearance. As mentioned in the previous section, the WISE data are 
less sensitive to photospheres at longer wavelengths, which is reflected in the 
limiting spectral types reached by the photospheric sequences.

In each diagram in Figure~\ref{fig:exc}, I have marked the threshold that I 
have used for identifying color excesses. If a star appears above the threshold 
for an excess in a given band but a detection in any band at a longer 
wavelength is consistent with a photosphere, an excess is not assigned to the 
first band. For the small fraction of the WISE sources ($\sim3$\%) that are 
separated by $>1\arcsec$ from a Gaia candidate, it was unclear whether the
spectral type and extinction of the latter could be adopted, so I identified 
excesses with diagrams like those in Figure~\ref{fig:exc} with W1 as a 
substitute for spectral type and no correction for extinction.

Table~\ref{tab:phot} contains three flags that 
indicate whether excesses were identified in W2, W3, and W4.
Flags are absent for non-detections. Eight sources are detected only in W1,
so they lack excess flags. Two of those objects do have disk classifications
from previous work that are based on Spitzer data.
I have previously used WISE to search for disks among some of
the sources in Table~\ref{tab:phot} \citep{luh12,esp18,esp20,luh20u,luh20lu}. 
A small number of those excess classifications have changed in this work,
primarily for data that are near the excess thresholds.

\subsection{Classification of Disks}

For each source from Table~\ref{tab:phot} that exhibits IR excess emission,
I have classified the evolutionary stage of its disk from among the following
options \citep{ken05,rie05,her07,luh10,esp12}:
{\it full disks} are optically thick at IR wavelengths and lack significant
clearing of primordial dust and gas; {\it transitional disks} have large 
inner holes in their dust components; {\it evolved disks} are becoming
optically thin at IR wavelengths but have not experienced significant clearing; 
{\it evolved transitional disks} are optically thin and have large inner holes; 
{\it debris disks} consist of second-generation dust produced by
collisions of planetesimals. All of these classes except for the latter
are considered primordial disks.
For reference, young stars can be classified based on the presence
or absence of a disk or infalling envelope \citep{lw84,lad87,and93,gre94}:
classes~0 and I (protostar with an infalling envelope and a primordial disk),
class~II (star with a primordial disk but no envelope), and class~III
(star that no longer has a primordial disk but that can have a debris disk).

\begin{figure}
\epsscale{1.4}
\plotone{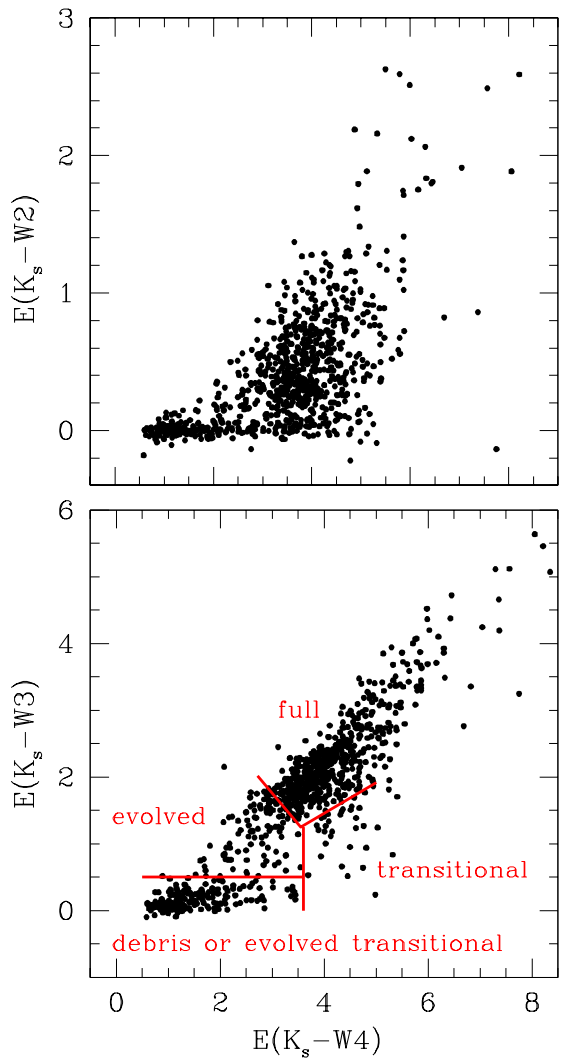}
\caption{
Extinction-corrected IR color excesses for the WISE
sources from Table~\ref{tab:phot} that have excesses.
The boundaries used for assigning disk classes are shown in the bottom diagram
(red solid lines).
}
\label{fig:exc2}
\end{figure}

I have assigned disk classes based on the sizes of the excesses in
$K_s-$W3 and $K_s-$W4 as done in my previous disk surveys with WISE
\citep{luh12,esp14,esp18}.
If a given WISE source is separated by $\leq1\arcsec$ from the nearest 2MASS
source, the value of $K_s$ from 2MASS is used for calculating those colors.
The color excesses, E($K_s-$W3) and E($K_s-$W4), are then calculated by 
subtracting the estimated reddening and the expected photospheric color 
for the spectral type in question \citep{luh21}.
Otherwise, if the WISE/2MASS separation is $>1\arcsec$, I have derived color 
excesses relative to W1, namely E(W1$-$W3) and E(W1$-$W4), which serve
as lower limits on E($K_s-$W3) and E($K_s-$W4) given that 
E($K_s-$W3)=E($K_s-$W1)+E(W1$-$W3) and E($K_s-$W4)=E($K_s-$W1)+E(W1$-$W4).
These limits should be close to the actual values for most sources
since most (94\%) disk-bearing members of Sco-Cen with $K_s$ data have 
E($K_s-$W1)$<$0.5.
The estimates of E($K_s-$W3) and E($K_s-$W4) 
are plotted in Figure~\ref{fig:exc2} with the criteria for the disk classes 
\citep{esp18}. As indicated in that diagram, the same criteria are used for 
debris and evolved transitional disks, which are indistinguishable in mid-IR 
data. If a source has a detection in W3 but not W4, I have classified it in the
following manner: full disk for E($K_s-$W3)$>$1.25, evolved or transitional
for E($K_s-$W3)=0.5--1.25, and debris or evolved transitional for 
E($K_s-$W3)$<$0.5. An object that has an excess in W2 and lacks detections
in W3 and W4 is listed as a full disk
since most disks that have W2 excesses are full \citep[e.g.,][]{luh12}.
I also have included a diagram with E($K_s-$W2) in Figure~\ref{fig:exc2}
to illustrate the sizes of the excesses in W2.

Sources that lack excesses in any of the WISE bands are omitted
from Figure~\ref{fig:exc2}.
Those stars are designated as class~III unless previous photometry from
Spitzer has detected excesses at wavelengths in which WISE lacks detections,
in which case the disk classifications from Spitzer are adopted
\citep[e.g.,][]{luh12,esp18}.
For instance, some sources have photospheric fluxes in W2 and W3 and lack
detections in W4 but have excesses detected by more sensitive Spitzer
data in a band similar to W4. 
A few sources are listed as Be stars based on previous work
\citep{hil69,hou75,irv90,riv13}.
Most protostars in Sco-Cen (which reside in Ophiuchus and Lupus) are
absent from Table~\ref{tab:phot} because
they are too faint at optical wavelengths for Gaia detections.
Some of the early-type stars have larger IR excesses than expected from
my adopted criteria for debris and evolved transitional disks but
have been previously classified as debris disks with unusually large
amounts of hot dust 
\citep{chen06,chen11,chen12,chen14,moo11,mel13,vic16}.
Those objects are listed as debris disks in my catalog.
All objects in Table~\ref{tab:phot} have disk classifications except for
six stars that lack detections in Spitzer or WISE bands longward of W1.

IR excesses are detected by WISE or Spitzer for 1340 of the 9675 sources in 
Table~\ref{tab:phot}, 1297 of which are within the Sco-Cen boundary from 
\citet{dez99}.
The latter consist of six Be stars and 822 full, 43 transitional, 124 evolved,
seven debris, 244 debris or evolved transitional, and 51 evolved or 
transitional disks.
Approximately 900 of these disks have been reported in previous studies
\citep{mam02,rie05,car06,car08,car09,all07,pad06,pad08,cie07,cie10,chen05,chen11,chen12,cha07,sch07,mer08,ria09,ria12,spe11,luh12,riz12,riz15,sch12,daw13,cot16,pec16,esp18,gol18,esp20,luh20u,tei20,luh20lu}.
The remaining $\sim$400 disks are newly identified in this work.
In \citet{luh21}, I used my census of Sco-Cen with Gaia EDR3
to examine the membership of the stars from some of the previous disk surveys.
Many of the samples contained mixtures of field stars and members of multiple 
Sco-Cen populations, which can have an impact on the interpretation of the
data in terms of disk statistics.
In addition, there are differences between this work and previous
surveys in terms of whether disks are detected for specific stars.
For instance, some of WISE data previously identified as excess emission
are flagged as false or unreliable detections in my catalog.
As a reminder, my census of disks is restricted to candidate
members of Sco-Cen identified with Gaia EDR3. Some of the previous disk surveys
include candidate members that lack parallax measurements from Gaia, such
as those with high extinction or very low masses.

\subsection{Candidates for Edge-on Disks}
\label{sec:edgeon}

\begin{figure}
\epsscale{1.2}
\plotone{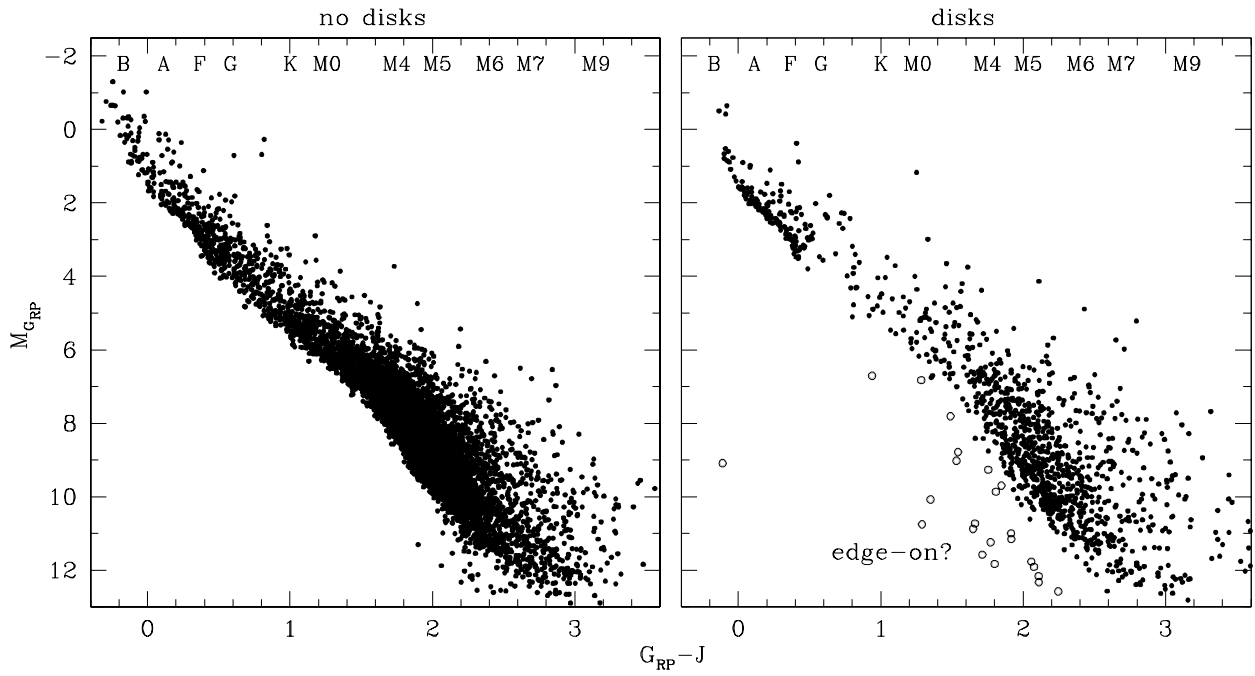}
\caption{
$M_{G_{\rm RP}}$ versus $G_{\rm RP}-J$ for candidate members of Sco-Cen that 
lack IR excesses from disks (left) and that do exhibit IR excesses (right).
Among the stars with disks, those that are unusually faint for their color
may be seen in scattered light due to an edge-on disk (open circles).
}
\label{fig:rj}
\end{figure}

Young stars that are occulted by edge-on disks are often observed
primarily in scattered light at optical and near-IR wavelengths.
As a result, they can appear unusually
faint at a given color or spectral type. I have attempted to use this signature
to identify candidates for edge-on disks among the sources in 
Table~\ref{tab:phot} that exhibit IR excesses.
I have selected a CMD constructed from
the reddest Gaia band, $G_{\rm RP}$, and $J$ because they
should contain the minimum amount of accretion-related emission (bright in
the UV) and dust emission (bright in the mid-IR) among the available bands
of photometry. In Figure~\ref{fig:rj}, I have plotted diagrams of
$M_{G_{\rm RP}}$ versus $G_{\rm RP}-J$ for WISE sources from 
Table~\ref{tab:phot} that are separated by $<1\arcsec$ from Gaia candidates
from \citet{luh21} that have $\sigma_{\pi}<0.5$~mas, renormalized unit weight
error (RUWE)$<1.6$, and positions within the boundary of Sco-Cen from
\citet{dez99}. Sources with and without IR excesses from disks are shown
in separate diagrams. The sequence of diskless stars exhibits a
well-defined lower boundary.
It is only at the faintest magnitudes that a few diskless stars appear
below the sequence formed by the bulk of the population.
Meanwhile, the disk-bearing sample contains a few dozen stars that are
unusually faint for their colors compared to the diskless stars, and
thus are candidates for edge-on disks. Those stars have ``edge-on?"
appended to their disk classifications in Table~\ref{tab:phot}.
Ten of the candidates have previous spectroscopy that confirms their youth 
\citep[][references therein]{luh21}, one of which is the star associated with 
HH~55. 
Many of the edge-on candidates are possible members of Upper Sco and
UCL/LCC \citep{luh21}, which have ages of $\sim11$ and 20~Myr, respectively.
Edge-on primordial disks at those relatively advanced ages
could be valuable for studies of disk structure.

\subsection{Estimate of Contamination among IR Excesses}

It is possible that some of the IR excesses detected toward candidate
members of Sco-Cen are due to very red background objects rather than
circumstellar disks. I have estimated the rate of such contamination
among excesses detected in W3 or W4.
The AllWISE Source Catalog contains $\sim$732,000 sources that have
W1$-$W3$>$1 or W1$-$W4$>$1, photometric errors less than 0.1~mag in these bands,
and positions within the boundary of Sco-Cen from \citet{dez99}.
Based on the corresponding surface density, $\sim8$ of the
sources in Table~\ref{tab:phot} are expected to have red contaminants
of that kind within angular separations of $3\arcsec$.
If a contaminant is more widely separated than that value, its centroid
in W4 is likely to be noticeably offset from the centroid of the Sco-Cen
candidate in W1 (Table~\ref{tab:pairs}).
The contamination estimate should be an upper limit since some of the
red background objects will not be sufficiently bright to produce an IR
excess when blended with the Sco-Cen candidates.
The presence of disks among the IR excess sources can be verified through
spectroscopy of accretion signatures and through high-resolution astrometry
of the long-wavelength components for comparison to the Gaia positions
of the short-wavelength components.

\section{Properties of the Disk Populations in Sco-Cen}

\subsection{Adopted Samples}
\label{sec:adopt}

In \citet{luh21}, to characterize the stellar populations in Sco-Cen,
I defined samples of candidate members that were meant to have minimal
contamination from field stars and between Sco-Cen populations.
I also use those samples of Gaia candidates in this work when characterizing
the disk populations in Sco-Cen. 
The samples consist of candidates 
with RUWE$<$1.6, $\sigma_{\pi}<0.5$~mas, and $\sigma_{BP}<0.1/\sigma_{RP}<0.1$
or $\sigma_G<0.1/\sigma_{RP}<0.1$ that have 1) Ophiuchus kinematics and a 
location within the boundary of Ophiuchus from \citet{esp18};
2) Upper Sco kinematics and a location outside of
Ophiuchus and within the triangular field from \citet{luh20u};
3) Lupus kinematics and a location within the fields toward clouds 1--4
from \citet{luh20lu}; 
4) V1062~Sco kinematics and a location within a radius
of $2\arcdeg$ from the center of that group;
and 5) UCL/LCC kinematics and a location within the boundary from \citet{dez99} 
and not within the fields for the other four samples.
For the disk analysis, I apply additional criteria to the
candidates: detections in W2, W3, or W4 are available so that an assessment
of IR excess was made; the WISE/Gaia separations are $<1\arcsec$;
the Gaia candidates are the closest matches for the WISE sources
(i.e., exclude a Gaia candidate if its best WISE match is closer to a
different Gaia source). In addition, the UCL/LCC candidates with
$l>340\arcdeg$ or $b<0\arcdeg$ are excluded since they appear to be slightly
younger than the remainder of that population \citep{luh21}.

\subsection{Comparison of Disk Populations within Sco-Cen}

Among the Gaia-selected candidate members, extinctions extend to higher
values in Ophiuchus than in any of the other Sco-Cen populations \citep{luh21}.
The Ophiuchus clouds also exhibit higher levels of extinction than the Lupus
clouds \citep{lom08}. 
As a result, a substantial fraction of Ophiuchus members are too reddened
for Gaia detections. For instance,
the census of Ophiuchus from \citet{esp20} contains $\sim90$
objects with spectral types of $\lesssim$M6 that lack Gaia parallaxes,
most of which have $A_V>5$.
The disk fraction is somewhat higher among those stars 
($\sim$2/3) than among the members in the Gaia-selected sample ($\sim$1/2).
In comparison, IR imaging of the Lupus clouds has identified only a few
candidate members that lack Gaia detections because of high extinction 
\citep{luh20lu}. The remaining populations in Sco-Cen are older and have
little obscuration \citep{luh21}, so the completeness of their disk fraction
samples is unaffected by extinction.

The Gaia-selected samples defined in the previous section have 
$N_{disk}/N_{total}=106/222$ (Ophiuchus), 62/93 (Lupus), 273/1098 (Upper Sco), 
23/337 (V1062~Sco), and 488/4889 (UCL/LCC) based on the data in 
Table~\ref{tab:phot}. The fractions decrease from
Ophiuchus/Lupus to Upper Sco to V1062~Sco/UCL/LCC, which is correlated
with increasing age, as expected. The disk fraction in the V1062~Sco group is
roughly similar to that in UCL/LCC, which is consistent with the fact
that they have similar ages \citep{luh20u,luh21}. Given the sizes of the
samples in V1062~Sco and UCL/LCC, the latter provides much better
statistical constraints on disk fractions at the age of the two populations.
Therefore, the V1062~Sco group is omitted from the remainder of the disk
analysis.

\begin{figure}
\epsscale{1.4}
\plotone{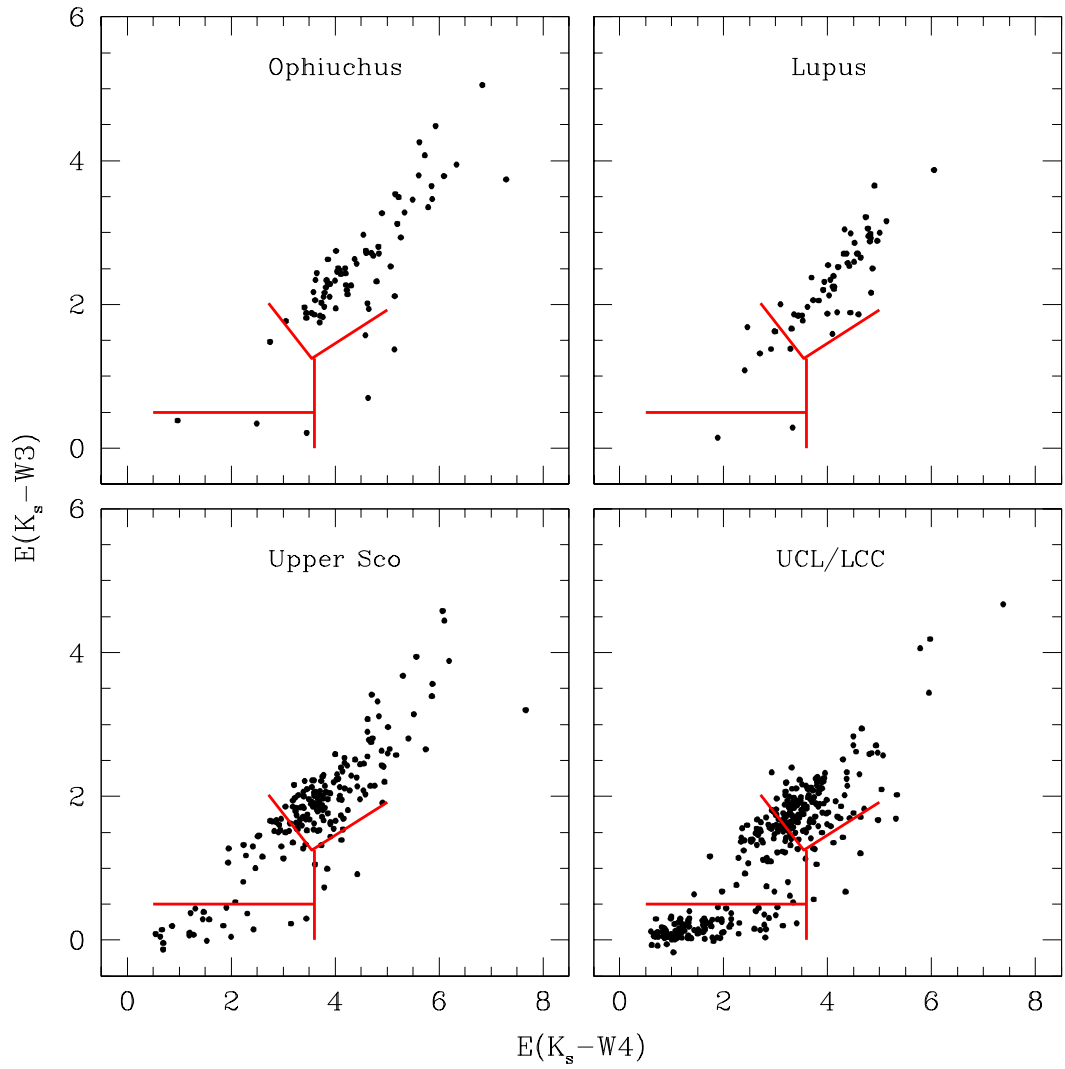}
\caption{
Extinction-corrected IR color excesses for candidate members of four
populations in Sco-Cen. The boundaries from Figure~\ref{fig:exc2}
used for assigning disk classes are indicated (red solid lines).
}
\label{fig:exc3}
\end{figure}

To further illustrate the evolution of the Sco-Cen disk populations with
age, I have plotted in Figure~\ref{fig:exc3} E($K_s-$W3) and E($K_s-$W4)
for the disk-bearing objects in the samples for Ophiuchus, Lupus, Upper Sco,
and UCL/LCC. 
As seen in previous disk surveys \citep{luh10,cesp14}, Figure~\ref{fig:exc3} 
shows that the youngest populations contain primarily full disks while disks 
in the more advanced stages become increasingly common in older regions.
Ophiuchus contains $\sim$50 previously proposed members
that exhibit IR excesses and are too obscured for Gaia detections, and hence
are absent from the sample shown in Figure~\ref{fig:exc3} \citep{eva09}.
Two objects of that kind are present in the Lupus clouds, consisting of
IRAS~15398$-$3359 and IRAS~16059$-$3857/Lupus~3~MMS \citep{mer08}.
Those embedded sources in Ophiuchus and Lupus,
many of which are protostellar, overlap with the full disks in
Figure~\ref{fig:exc3} and extend to redder colors.

\subsection{Excess and Disk Fractions for Upper Sco and UCL/LCC}

Because Upper Sco and UCL/LCC contain a large majority of the stars in Sco-Cen,
they are the best populations in Sco-Cen for measuring excess and disk
fractions as a function of spectral type, which serves as a proxy for stellar
mass. Measurements of that kind have been performed with previous samples of
candidates in Upper Sco using photometry from Spitzer and WISE
\citep{car06,car09,luh12,luh20u}.
The most recent study, \citet{luh20u}, considered a sample of candidates
that extended across a large field toward Upper Sco and that included
objects that had been vetted with Gaia DR2 data and sources that lacked
Gaia parallaxes (e.g., brown dwarfs).
In this work, my sample is restricted to candidates selected from Gaia EDR3
within a smaller field that encompasses the central concentration of Upper Sco
members (Section~\ref{sec:adopt}) in order to minimize contamination from
UCL/LCC, which overlaps spatially and kinematically with Upper Sco 
\citep{luh21}. As a result, the sample from \citet{luh20u} provides better
statistical constraints on excess fractions (particularly at early types)
and reaches later spectral types while the sample in this work should
have better reliability.

For the Upper Sco and UCL/LCC samples defined in Section~\ref{sec:adopt},
I have calculated excess fractions in W2, W3, and W4 for up to eight ranges
of spectral type between B0 and M6. All sources in these samples have 
detections in W2 but some are not detected in W3 and W4.
For the latter bands, the excess fractions are reported only for ranges
of spectral type in which most candidates have detections so that the
measurements are not biased against the faintest stars at given spectral type, 
which tend to be those with little or no excess. 
As done in \citet{luh12} and \citet{luh20u}, I calculate one excess fraction 
for three classes of primordial disks (full, evolved, transitional) and
one for debris and evolved transitional disks. The latter two classes are
combined because they are indistinguishable in the available mid-IR data.
Be stars are included among the sources that lack excesses since their IR 
emission does not arise from primordial or debris disks \citep{riv13}.
The statistical errors in the excess fractions are taken from \citet{geh86}.

If some of the candidate members of Upper Sco and UCL/LCC are field stars, 
the excess fractions would be underestimated
since field stars are unlikely to exhibit IR excesses.
In \citet{luh21}, I estimated the amount of field star contamination
among the candidate members of Sco-Cen considered in this work.
Among the spectral type ranges in which
excess fractions are calculated, the highest contamination should occur at
F4--G2, where $\sim5$ of the 89 UCL/LCC candidates are expected to be
field stars. The estimated contamination is lower ($<2$\%) in all other
spectral type ranges for UCL/LCC. The contamination should be much lower
in the Upper Sco sample due to the small size of the field in which
candidates were selected. Thus, field star contamination may lead to
slightly underestimated excess fractions in UCL/LCC at F4--G2 and should
have negligible effects on all other measurements.

\begin{figure}
\epsscale{1.4}
\plotone{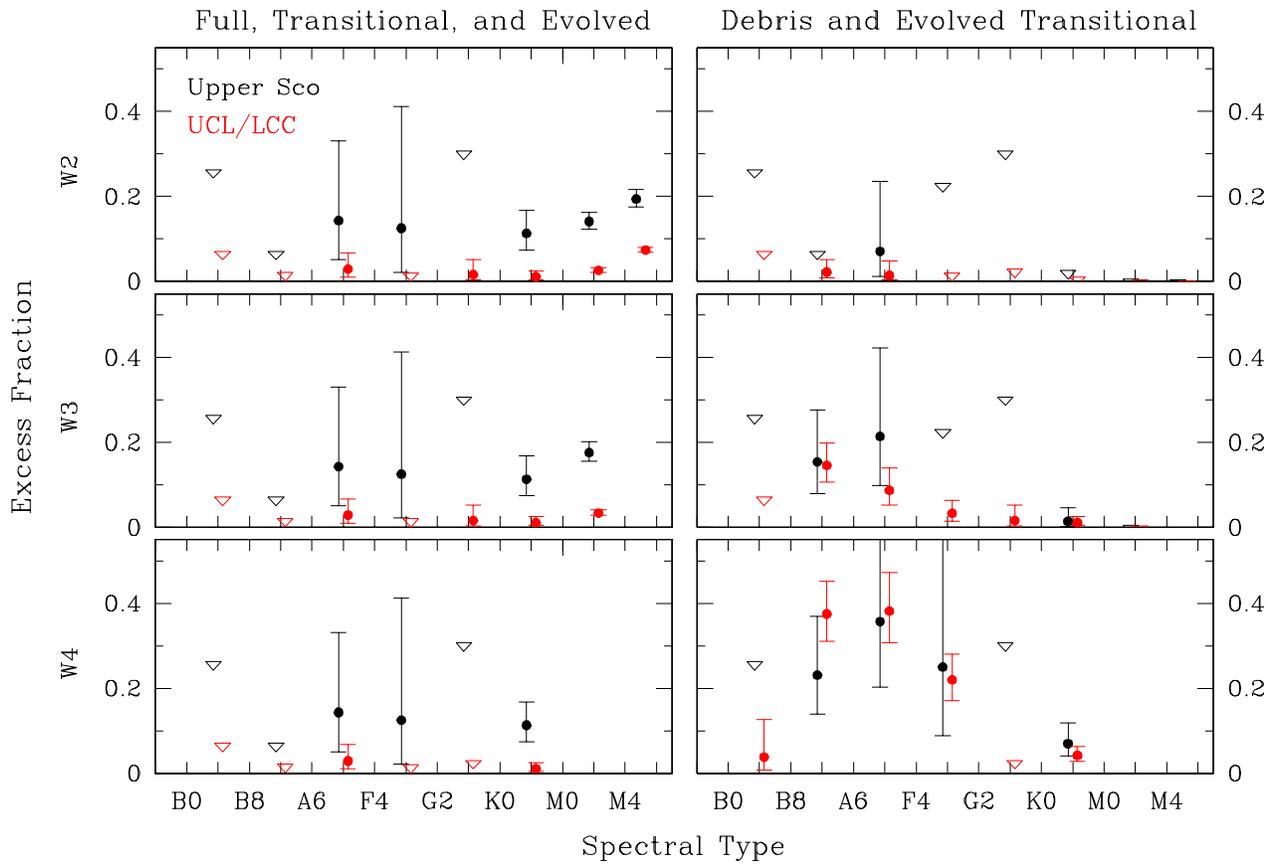}
\caption{
Excess fractions versus spectral type in Upper Sco and UCL/LCC
for full, transitional, and evolved disks (left) and debris and evolved
transitional disks (right, Tables~\ref{tab:fexu} and \ref{tab:fexc}).  
For each band, data are shown only down to the latest spectral type at which 
most candidate members are detected. The data for Upper Sco and 
UCL/LCC are offset slightly left and right, respectively, to avoid overlap.
The triangles represent 1~$\sigma$ upper limits.
}
\label{fig:fex}
\end{figure}

In Tables~\ref{tab:fexu} and \ref{tab:fexc} and Figure~\ref{fig:fex},
I present the excess fractions in W2, W3, and W4 as a function of spectral
type for the samples of candidate members of Upper Sco and UCL/LCC
defined in Section~\ref{sec:adopt}. As in Section~\ref{sec:exc}, photometric
estimates of spectral types are used when spectroscopic data are not available.
The measurements for Upper Sco are qualitatively similar to those for
the Upper Sco sample in \citet{luh20u} except that the statistical errors
for the early-type stars are larger in this study because of the smaller
field. The most notable feature of the Upper Sco data is the increase
in excess fractions with later spectral types, which was first detected
by \citet{car06}. That trend continues beyond the latest spectral type
considered here, M6 ($\sim0.1$~$M_\odot$), down to early L 
($\sim0.01$~$M_\odot$) \citep{luh12,luh20u}. The debris disk fraction in 
Upper Sco peaks at A and F types, as discussed in previous work 
\citep{car09,chen11}.

The excess fractions for UCL/LCC represent a significant improvement over
previous data in that population because the sample of candidates
is much larger and more reliable than those in previous studies,
particularly at later spectral types. Like Upper Sco, UCL/LCC exhibits
increasing excess fractions with later spectral types and high debris disk
fractions among A and F stars. In addition to appearing in the diagram
of excess fractions in Figure~\ref{fig:fex}, the sharp drop in 
debris disk fractions from F to G types is also evident in the
CMD of disk-bearing stars in Figure~\ref{fig:rj}.
Meanwhile, the excess fractions of primordial disks are lower in UCL/LCC than
in Upper Sco at all spectral types, as expected given the difference in ages.
The drop in those fractions from Upper Sco to UCL/LCC is larger at
earlier types; the ratio of their fractions in W2 is $\sim$9.7, 6.1, and 3.0
at K0--M0, M0--M4, and M4--M6, respectively. Upper Sco had previously
demonstrated that disks are dispersing more quickly at earlier types by
the age $\sim11$~Myr, and UCL/LCC shows that the trend continues between
$\sim$11 and 20~Myr.

\begin{figure}
\epsscale{1.4}
\plotone{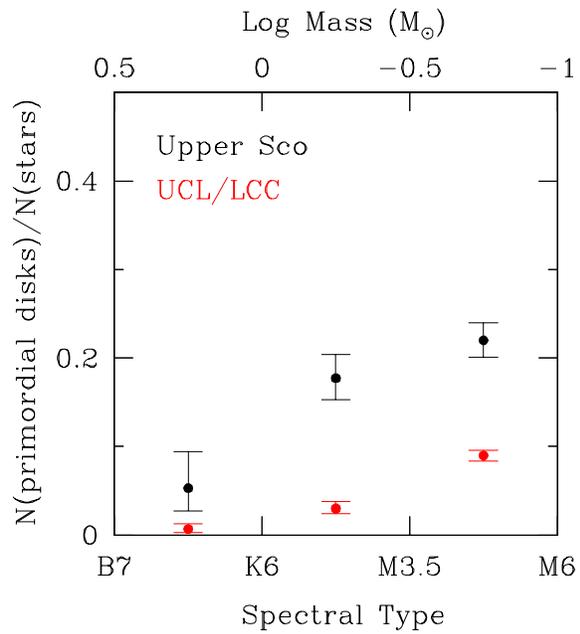}
\caption{
Fractions of candidate members of Upper Sco and UCL/LCC with primordial disks
(full, transitional, evolved) as a function of spectral type 
(Table~\ref{tab:frac}).
The boundaries of the spectral type bins were chosen to correspond
approximately to logarithmic intervals of mass.
}
\label{fig:frac}
\end{figure}

In addition to the excess fractions, I have calculated the primordial
disk fractions of the candidates in the Upper Sco and UCL/LCC samples for 
three ranges of spectral types that correspond roughly to logarithmic
intervals of stellar mass \citep{bar98,bar15}.
The numerator in the disk fraction consists of
full, evolved, and transitional disks, or all class~II objects except
evolved transitional disks. The disk fractions are presented
in Table~\ref{tab:frac} and Figure~\ref{fig:frac}. The fractions are
lower in UCL/LCC than in Upper Sco by factors of $\sim$10, 5.7, and 2.5
at B7--K5.5, K6--M3.5, and M3.75--M6, respectively.

\subsection{Relative Ages of Stars with and without Disks}

The high-precision parallaxes and photometry from Gaia have facilitated
measurements of relative ages of populations in Sco-Cen via their sequences
of low-mass stars in the Hertzsprung-Russell (H-R) diagram
\citep[e.g.,][]{dam19,luh20u}.
I have used that method to constrain the relative ages of the disk-bearing
and diskless stars in Upper Sco and UCL/LCC.
$M_{G_{\rm RP}}$ is selected for the vertical axis of the H-R diagram since
$G_{\rm RP}$ is less affected by extinction and accretion-related emission
than the other two Gaia bands. 
The diagrams are constructed with colors instead of spectral types
(i.e., CMDs) since most low-mass stars in UCL/LCC lack spectroscopy.
$G_{\rm RP}-J$ is used for one of the CMDs as done in
Figure~\ref{fig:rj} (Section~\ref{sec:edgeon}). I also employ a CMD
with $G_{\rm BP}-G_{\rm RP}$ because both bands have high precision,
although some of the disk-bearing stars may have noticeable excesses in
$G_{\rm BP}$ due to accretion.
I consider the samples of candidate members of Upper Sco and UCL/LCC defined
in Section~\ref{sec:adopt} with the additional criteria of 
$\sigma_{\pi}<0.3$~mas and $A_K<0.1$. A smaller threshold on parallax
error is adopted since the relative ages rely on the precisions of the
absolute magnitudes. Only stars with low extinction are included to minimize
errors in the relative ages due to errors in the extinction corrections.
The photometry for the CMDs has been corrected for extinction using the
estimates from \citet{luh21}.

\begin{figure}
\epsscale{1.2}
\plotone{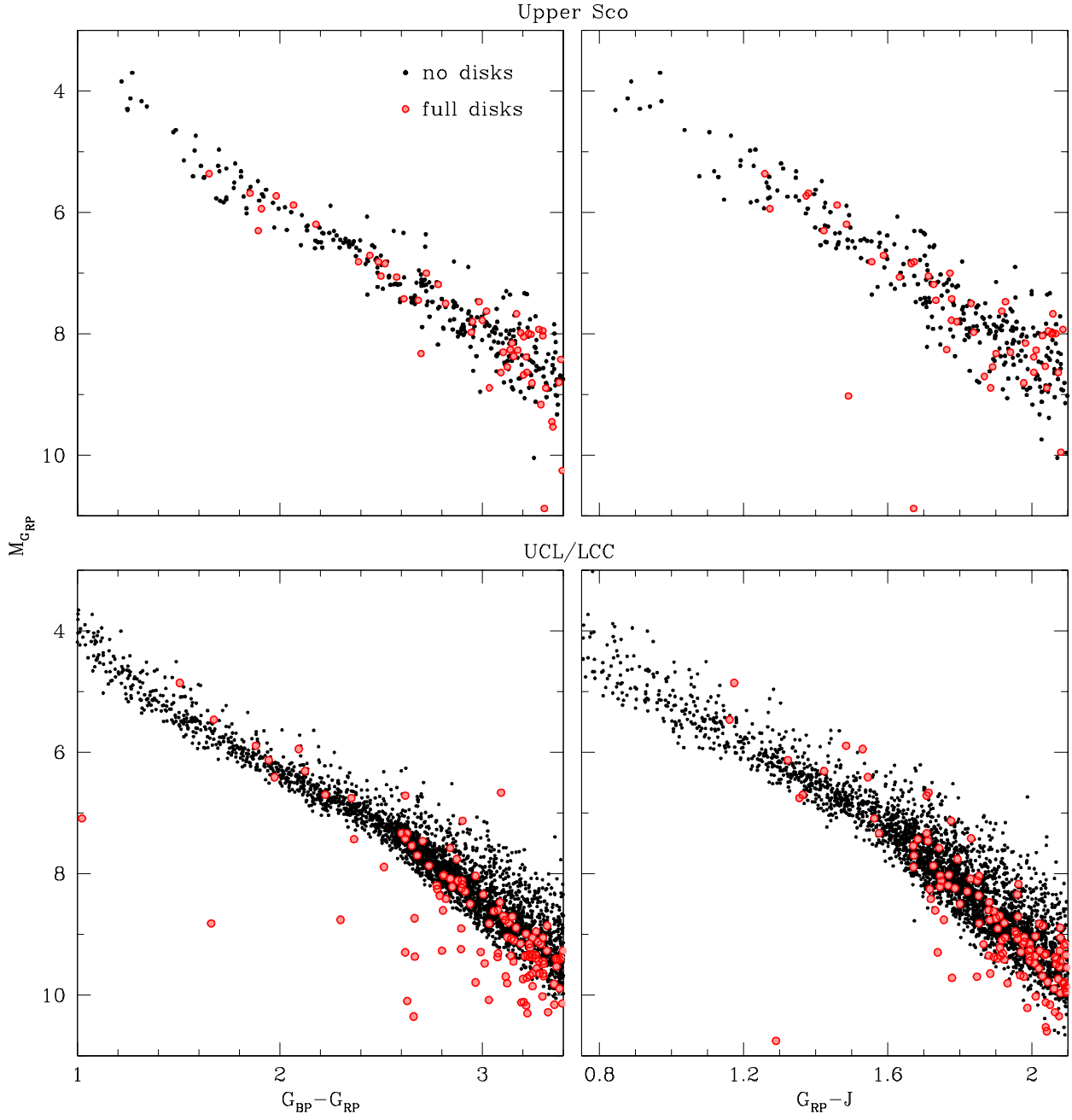}
\caption{
$M_{G_{\rm RP}}$ versus $G_{\rm BP}-G_{\rm RP}$ and $G_{\rm RP}-J$ for
candidate members of Upper Sco and UCL/LCC that lack disks (black points) and
that have full disks (red points).
}
\label{fig:ages}
\end{figure}

In Figure~\ref{fig:ages}, I have plotted CMDs for the low-mass stars in
Upper Sco and UCL/LCC that have been classified as class~III or
full disks in Table~\ref{tab:phot}. Only the earliest and latest evolutionary
stages are shown in order to facilitate the detection of systematic differences
in ages among the stages. For each CMD, I calculated the offsets in 
$M_{G_{\rm RP}}$ of the late-type stars ($G_{\rm BP}-G_{\rm RP}=1.6$--3.4, 
$G_{\rm RP}-J=1.14$--2.1) from the median sequence for
the combined sample of full disks and class III objects. Stars appearing
below the sequence in a CMD were excluded from the analysis, most of which
are disk-bearing stars and likely have accretion-related emission
in $G_{\rm BP}$ or edge-on disks (Section~\ref{sec:edgeon}).
I then calculated the difference between the median offsets for the two
evolutionary stages for a given CMD and population. 
The error in the difference was characterized using the median absolute
deviation for the distribution of differences produced by bootstrapping.
The resulting differences in $M_{G_{\rm RP}}$ from the $G_{\rm BP}-G_{\rm RP}$ 
and $G_{\rm RP}-J$ CMDs are $-0.02\pm0.05$ and $-0.01\pm0.09$~mag for Upper Sco
and $0.05\pm0.06$ and $0.04\pm0.05$~mag for UCL/LCC where positive values
correspond to older ages for the full disks. Thus, I do not find any
significant differences in the ages of full disks and class III objects
in Upper Sco or UCL/LCC. For perspective, a difference of 0.1~mag in
luminosity is predicted to correspond to $\sim$0.06~dex in age for low-mass 
stars at 10--20 Myr \citep{bar15,cho16,dot16,fei16}.

\section{Conclusions}

I recently used data from Gaia EDR3 to identify candidate members
of the stellar populations within the Sco-Cen complex \citep{luh21},
which consist of Ophiuchus, Lupus, Upper Sco, the V1062~Sco group, and UCL/LCC.
In this study, I have performed a survey for circumstellar disks among
those candidates using mid-IR photometry from WISE.
The results are summarized as follows:

\begin{enumerate}

\item
In \citet{luh21}, I used astrometry and photometry from Gaia EDR3 to
identify 10,509 candidate members of Sco-Cen. I also presented 195 additional
sources that did not satisfy the kinematic criteria for membership 
but that are possible companions to candidates in the first sample.
The tabulations of those candidates in \citet{luh21} included the
names of the closest 2MASS and WISE sources within $3\arcsec$. 
The 9675 matching WISE sources are compiled in Table~\ref{tab:phot}, which
contains their photometry from 2MASS and WISE.
I have inspected the WISE images for all objects and have flagged
detections that appear to be false or unreliable.

\item
I have used the WISE colors to identify sources in Table~\ref{tab:phot}
that exhibit IR excesses from disks and I have classified the
evolutionary stages of the detected disks using the sizes of the excesses.
IR excesses are detected for more than 1200 of the WISE sources that are
within the Sco-Cen boundary from \citet{dez99}, $\sim400$ of which
are newly detected in this work. 
Because of its proximity and size, UCL/LCC offers the largest available sample
of disks ($>$500) for any population near its age ($\sim$20~Myr).
Among the disk-bearing objects in Sco-Cen, I have identified 23 candidates
for edge-on disks based on their unusually faint positions in a CMD.

\item
For the two largest stellar populations in Sco-Cen, Upper Sco and UCL/LCC,
I have measured the fractions of sources that have 
excesses in W2, W3, and W4 as a function of spectral type. Separate excess 
fractions are reported for disks that are full, transitional, or evolved and 
disks that are evolved transitional or debris. In addition, I have measured
the fractions of sources with disks that are full, transitional, or evolved
(i.e., class~II objects with the exception of evolved transitional disks).
The new excess and disk fractions in Upper Sco are qualitatively similar to
measurements for earlier samples of candidate members
\citep{car06,luh12,luh20u}. As found in the previous work,
the most notable feature of the data in Upper Sco is an increase in disk
fractions with later spectral types, which indicates that disk lifetimes
are longer for stars at lower masses.

\item
The new census of stars and disks in UCL/LCC ($\sim20$~Myr) has provided the 
tightest statistical constraints to date on the disk fractions of low-mass
stars for any single age beyond that of Upper Sco ($\sim11$~Myr).
As in Upper Sco, the disk fraction in UCL/LCC is higher at later spectral
types. That trend is more pronounced in UCL/LCC, where the fractions are 
lower than those in Upper Sco by factors of $\sim$10, 5.7,
and 2.5 at B7--K5.5, K6--M3.5, and M3.75--M6, respectively.
The data in UCL/LCC also demonstrate that the disk fraction for low-mass stars
remains non-negligible at an age of 20~Myr (0.09$\pm$0.01).

\item
I have used CMDs to constrain the relative ages of disk-bearing and diskless
low-mass stars in Upper Sco and UCL/LCC. No significant differences in
ages are found at a level of $\gtrsim0.06$~dex.

\end{enumerate}

\acknowledgements

This work used data from the European Space Agency (ESA)
mission Gaia (\url{https://www.cosmos.esa.int/gaia}), processed by
the Gaia Data Processing and Analysis Consortium (DPAC,
\url{https://www.cosmos.esa.int/web/gaia/dpac/consortium}). Funding
for the DPAC has been provided by national institutions, in particular
the institutions participating in the Gaia Multilateral Agreement.
2MASS is a joint project of the University of Massachusetts and IPAC
at Caltech, funded by NASA and the NSF.
WISE is a joint project of the University of California, Los Angeles,
and the JPL/Caltech, funded by NASA.
This work used data from the Spitzer Space Telescope and the
NASA/IPAC Infrared Science Archive, operated by JPL under contract
with NASA, and the VizieR catalog access tool and the SIMBAD database, both
operated at CDS, Strasbourg, France.
The Center for Exoplanets and Habitable Worlds is supported by the
Pennsylvania State University, the Eberly College of Science, and the
Pennsylvania Space Grant Consortium.

\clearpage

\begin{deluxetable}{ll}
\tabletypesize{\scriptsize}
\tablewidth{0pt}
\tablecaption{IR Photometry and Disk Classifications for Candidate Members
of Sco-Cen\label{tab:phot}}
\tablehead{
\colhead{Column Label} &
\colhead{Description}}
\startdata
wise & Source name from WISE catalogs\tablenotemark{a} \\
RAdeg & Right ascension (J2000) from WISE\\
DEdeg & Declination (J2000) from WISE\\
2m & Closest 2MASS source within $3\arcsec$\\
2msep & Angular separation between WISE and 2MASS\\
Jmag & 2MASS $J$ magnitude \\
e\_Jmag & Error in Jmag \\
Hmag & 2MASS $H$ magnitude \\
e\_Hmag & Error in Hmag \\
Ksmag & 2MASS $K_s$ magnitude \\
e\_Ksmag & Error in Ksmag \\
W1mag & WISE W1 magnitude \\
e\_W1mag & Error in W1mag \\
f\_W1mag & Flag on W1mag\tablenotemark{b} \\
W2mag & WISE W2 magnitude \\
e\_W2mag & Error in W2mag \\
f\_W2mag & Flag on W2mag\tablenotemark{b} \\
W3mag & WISE W3 magnitude \\
e\_W3mag & Error in W3mag \\
f\_W3mag & Flag on W3mag\tablenotemark{b} \\
W4mag & WISE W4 magnitude \\
e\_W4mag & Error in W4mag \\
f\_W4mag & Flag on W4mag\tablenotemark{b} \\
ExcW2 & Excess present in W2? \\
ExcW3 & Excess present in W3? \\
ExcW4 & Excess present in W4? \\
DiskType & Disk Type\tablenotemark{c}
\enddata
\tablecomments{This table is available in its entirety in a machine-readable form.}
\tablenotetext{a}{Source name from AllWISE Source Catalog, AllWISE Reject
Catalog, or WISE All-Sky Source Catalog.}
\tablenotetext{b}{nodet = nondetection; false = detection from
WISE catalog appears to be false or unreliable based on visual inspection.}
\tablenotetext{c}{Candidates for edge-on disks from Figure~\ref{fig:rj}
have ``edge-on?" appended to their disk classifications.}
\end{deluxetable}

\clearpage

\begin{deluxetable}{llll}
\tabletypesize{\scriptsize}
\tablewidth{0pt}
\tablecaption{Pairs Partially Resolved by WISE\label{tab:pairs}}
\tablehead{
\colhead{WISE Source} &
\colhead{Dominates at W1\tablenotemark{a}} &
\colhead{Dominates at W3 or W4\tablenotemark{a}} &
\colhead{Possible}\\
\colhead{} &
\colhead{} &
\colhead{} &
\colhead{Binary?}}
\startdata
WISEA J122420.62$-$544354.1 & 6076080033122502400/6076080033122502528 & 6076080033106941440 & no \\
WISEA J132335.83$-$471846.9 & 6083750638577673088 & 6083750638540951552 & yes \\
WISEA J133255.99$-$580710.8 & 5870395561188264320 & 5870395561176603392 & no \\
WISEA J134906.74$-$441317.8 & 6108605060227272960 & 6108605064523357056 & yes \\
WISEA J140659.03$-$423842.8 & 6109774841820067968 & 6109774841820067072 & yes \\
WISEA J144449.43$-$435008.5 & 6100428065273180672 & 6100428095333967744 & \nodata \\
WISEA J145833.85$-$422342.9 & 5908685327718523520 & 5908685332019899520 & \nodata \\
WISEA J152759.91$-$320804.8 & 6206980427151256192 & 6206980431449776640 & yes \\
WISEA J153102.01$-$341635.0 & 6014310603731467008/6014310638092127616 & 6014310638092127616 & yes \\
WISEA J153639.94$-$342143.2 & 6014168251335936640 & 6014168251336833024 & \nodata \\
WISEA J154335.49$-$384805.2 & 6009110841455006336 & 6009110807095267712 & yes \\
WISEA J155548.78$-$251224.1 & 6235742349962813824 & 6235742349962814592 & yes \\
WISEA J155734.29$-$232112.4 & 6237142264484167296 & 6237142268777708928 & yes? \\
WISEA J160408.41$-$190730.2 & 6247323329838233472/6247323334141396992 & 6247323334141396992 & yes \\
WISEA J160842.76$-$390617.6 & 5997082214303850880 & 5997082218616859264 & yes \\
WISEA J161002.77$-$234440.5 & 6242136319309548928/6242137796783514240 & 6242136319309548928 & yes \\
WISEA J161019.15$-$250230.6 & 6049748791208799488 & 6049748786908497408 & yes \\
WISEA J161343.96$-$373644.3 & 5997644343928854656/5997644343928856704 & 5997644343928854656 & yes \\
WISEA J162459.13$-$252118.2 & 6048951366100474368 & non-Gaia source & \nodata \\
WISE J162902.95$-$242749.3 & 6049102587605960064 & 6049102621963483136 & yes \\
WISEA J183123.23$-$344500.5 & 6734855001900424704 & non-Gaia source & \nodata 
\enddata
\tablenotetext{a}{Designations are from Gaia EDR3.}
\end{deluxetable}

\clearpage

\begin{deluxetable}{lllll}
\tablecolumns{5}
\tabletypesize{\scriptsize}
\tablewidth{0pt}
\tablecaption{Excess Fractions in Upper Sco\label{tab:fexu}}
\tablehead{
\colhead{Spectral Type} &
\colhead{Mass\tablenotemark{a}} &
\colhead{W2} &
\colhead{W3} &
\colhead{W4} \\
\colhead{} &
\colhead{($M_\odot$)} &
\colhead{} &
\colhead{} &
\colhead{}
}
\startdata
\cutinhead{Full, Transitional, and Evolved Disks}
B0--B8 & 3--18 & 0/7=$<$0.26 & 0/7=$<$0.26 & 0/7=$<$0.26\\
B8--A6 & 1.8--3 & 0/26=$<$0.07 & 0/26=$<$0.07 & 0/26=$<$0.07\\
A6--F4 & 1.5--1.8 & 2/14=$0.14^{+0.19}_{-0.09}$ & 2/14=$0.14^{+0.19}_{-0.09}$ & 2/14=$0.14^{+0.19}_{-0.09}$\\
F4--G2 & 1.4--1.5 & 1/8=$0.12^{+0.29}_{-0.10}$ & 1/8=$0.12^{+0.29}_{-0.10}$ & 1/8=$0.12^{+0.29}_{-0.10}$\\
G2--K0 & 1.3--1.4 & 0/6=$<$0.31 & 0/6=$<$0.31 & 0/6=$<$0.31\\
K0--M0 & 0.7--1.3 & 8/71=$0.11^{+0.06}_{-0.04}$ & 8/71=$0.11^{+0.06}_{-0.04}$ & 8/71=$0.11^{+0.06}_{-0.04}$\\
M0--M4 & 0.2--0.7 & 53/376=$0.14\pm0.02$ & 66/374=$0.18\pm0.02$ & \nodata\\
M4--M6 & 0.1--0.2 & 100/515=$0.19\pm0.02$ & \nodata & \nodata\\
\cutinhead{Debris and Evolved Transitional Disks}
B0--B8 & 3--18 & 0/7=$<$0.26 & 0/7=$<$0.26 & 0/7=$<$0.26\\
B8--A6 & 1.8--3 & 0/26=$<$0.07 & 4/26=$0.15^{+0.12}_{-0.07}$ & 6/26=$0.23^{+0.14}_{-0.09}$\\
A6--F4 & 1.5--1.8 & 1/14=$0.07^{+0.16}_{-0.06}$ & 3/14=$0.21^{+0.21}_{-0.12}$ & 5/14=$0.36^{+0.24}_{-0.15}$\\
F4--G2 & 1.4--1.5 & 0/8=$<$0.23 & 0/8=$<$0.23 & 2/8=$0.25^{+0.33}_{-0.16}$\\
G2--K0 & 1.3--1.4 & 0/6=$<$0.31 & 0/6=$<$0.31 & 0/6=$<$0.31\\
K0--M0 & 0.7--1.3 & 0/71=$<$0.03 & 1/71=$0.01^{+0.03}_{-0.01}$ & 5/71=$0.07^{+0.05}_{-0.03}$\\
M0--M4 & 0.2--0.7 & 0/376=$<$0.01 & 0/374=$<$0.01 & \nodata\\
M4--M6 & 0.1--0.2 & 0/515=$<$0.004 & \nodata & \nodata\\
\enddata
\tablenotetext{a}{Masses that correspond to the given range of spectral types
for an age of 10~Myr \citep{bar98,bar15,cho16,dot16}.}
\end{deluxetable}

\begin{deluxetable}{lllll}
\tablecolumns{5}
\tabletypesize{\scriptsize}
\tablewidth{0pt}
\tablecaption{Excess Fractions in UCL/LCC\label{tab:fexc}}
\tablehead{
\colhead{Spectral Type} &
\colhead{Mass\tablenotemark{a}} &
\colhead{W2} &
\colhead{W3} &
\colhead{W4} \\
\colhead{} &
\colhead{($M_\odot$)} &
\colhead{} &
\colhead{} &
\colhead{}
}
\startdata
\cutinhead{Full, Transitional, and Evolved Disks}
B0--B8 & 3--11 & 0/26=$<$0.07 & 0/26=$<$0.07 & 0/26=$<$0.07\\
B8--A6 & 1.7--3 & 0/89=$<$0.02 & 0/89=$<$0.02 & 0/88=$<$0.02\\
A6--F4 & 1.4--1.7 & 2/69=$0.03^{+0.04}_{-0.02}$ & 2/69=$0.03^{+0.04}_{-0.02}$ & 2/68=$0.03^{+0.04}_{-0.02}$\\
F4--G2 & 1.1--1.4 & 0/92=$<$0.02 & 0/92=$<$0.02 & 0/91=$<$0.02\\
G2--K0 & 1.0--1.1 & 1/64=$0.02^{+0.04}_{-0.01}$ & 1/64=$0.02^{+0.04}_{-0.01}$ & 0/62=$<$0.03\\
K0--M0 & 0.7--1.0 & 2/189=$0.01\pm0.01$ & 2/189=$0.01\pm0.01$ & 2/189=$0.01\pm0.01$\\
M0--M4 & 0.2--0.7 & 26/1005=$0.03\pm0.01$ & 34/1004=$0.03\pm0.01$ & \nodata\\
M4--M6 & 0.1--0.2 & 159/2149=$0.07\pm0.01$ & \nodata & \nodata\\
\cutinhead{Debris and Evolved Transitional Disks}
B0--B8 & 3--11 & 0/26=$<$0.07 & 0/26=$<$0.07 & 1/26=$0.04^{+0.09}_{-0.03}$\\
B8--A6 & 1.7--3 & 2/89=$0.02^{+0.03}_{-0.01}$ & 13/89=$0.15^{+0.05}_{-0.04}$ & 33/88=$0.38^{+0.08}_{-0.07}$\\
A6--F4 & 1.4--1.7 & 1/69=$0.01^{+0.03}_{-0.01}$ & 6/69=$0.09^{+0.05}_{-0.03}$ & 26/68=$0.38^{+0.09}_{-0.07}$\\
F4--G2 & 1.1--1.4 & 0/92=$<$0.02 & 3/92=$0.03^{+0.03}_{-0.02}$ & 20/91=$0.22^{+0.06}_{-0.05}$\\
G2--K0 & 1.0--1.1 & 0/64=$<$0.03 & 1/64=$0.02^{+0.04}_{-0.01}$ & 0/62=$<$0.03\\
K0--M0 & 0.7--1.0 & 0/189=$<$0.01 & 2/189=$0.01\pm0.01$ & 8/189=$0.04^{+0.02}_{-0.01}$\\
M0--M4 & 0.2--0.7 & 1/1005=$0.001^{+0.002}_{-0.001}$ & 1/1004=$0.001^{+0.002}_{-0.001}$ & \nodata\\
M4--M6 & 0.1--0.2 & 0/2149=$<$0.001 & \nodata & \nodata\\
\enddata
\tablenotetext{a}{Masses that correspond to the given range of spectral types
for an age of 20~Myr \citep{bar98,bar15,cho16,dot16}.}
\end{deluxetable}

\clearpage

\begin{deluxetable}{ll}
\tabletypesize{\scriptsize}
\tablewidth{0pt}
\tablecaption{Disk Fractions\label{tab:frac}}
\tablehead{
\colhead{Spectral Type} &
\colhead{N(primordial disks)/N(stars)}}
\startdata
\cutinhead{Upper Sco}
B7--K5.5 & 4/76=$0.05^{+0.04}_{-0.03}$\\
K6--M3.5 & 55/311=$0.18^{+0.03}_{-0.02}$\\
M3.75--M6 & 139/633=$0.22\pm0.02$\\
\cutinhead{UCL/LCC}
B7--K5.5 & 3/452=$0.007^{+0.006}_{-0.004}$\\
K6--M3.5 & 22/725=$0.03\pm0.01$\\
M3.75--M6 & 222/2488=$0.09\pm0.01$\\
\enddata
\end{deluxetable}

\clearpage

\end{document}